\documentclass[10pt,a4paper]{iopart}
\usepackage{epsf}

\begin{document}
\jl{6}
\title[Photon capture cones and embedding diagrams of the Ernst spacetime]%
  {Photon capture cones and embedding diagrams\\ of the Ernst spacetime%
  \footnote{Published in Class.\ Quantum Grav.\ \textbf{16} (1999),
    pp.~1377--1387.}}
\author{Z Stuchl\'{\i}k\footnote{E-mail address: Zdenek.Stuchlik@fpf.slu.cz}
  and S Hled\'{\i}k\footnote{E-mail address: Stanislav.Hledik@fpf.slu.cz}}

%  Abdus Salam International Centre for Theoretical Physics,
%  Strada Costiera~11, 340~14~Trieste, Italy

\address{Institute of Physics, Faculty of Philosophy and Science, Silesian
  University in~Opava, Bezru\v{c}ovo n\'am.~13, CZ-746\,01~Opava,
  Czech Republic}

\begin{abstract}
The differences between the character of the Schwarzschild and Ernst
spacetimes are illustrated by comparing the photon capture cones, and the
embedding diagrams of the $t={\rm const}$ sections of the equatorial planes of both
the ordinary and optical reference geometry of these spacetimes.  The non-flat
asymptotic character of the Ernst spacetime reflects itself in two manifest
facts: the escape photon cones correspond to purely outward radial direction,
and the embedding diagrams of both the ordinary and optical geometry shrink to
zero radius asymptotically. Using the properties of the embedding diagrams,
regions of these spacetimes which could have similar character are estimated,
and it is argued that they can exist for the Ernst spacetimes with a
sufficiently low strength of the magnetic field.
\end{abstract}

  % Exact solutions

\maketitle

%%%% KEYSTROKE-SAVING MACROS (quite harmless :-) %%%%
\def\be{\begin{equation}}
\def\ee{\end{equation}}
\def\bea{\begin{eqnarray}}
\def\eea{\end{eqnarray}}
\def\expon#1{{\rm e}^{#1}}
\def\d{{\rm d}}
\def\D{{\rm D}}
\def\pder#1#2{\frac{\partial#1}{\partial#2}}
\def\oder#1#2{\frac{\d#1}{\d#2}}
\def\fcomma{{}}
\arraycolsep 0.28em

\section{Introduction}  \label{intro}

The static Ernst spacetime~\cite{Er} is the only exact solution of Einstein's
equations known to represent the spacetime of a spherically symmetric massive
body or black hole of mass $M$ immersed in an otherwise homogeneous magnetic
field. If the magnetic field disappears, the geometry simplifies to the
Schwarzschild geometry. Therefore, sometimes the Ernst spacetime is called
magnetized Schwarzschild spacetime. Usually it is believed that for extended
structures like galaxies both the effects of general relativity and the role
of a magnetic field can be ignored. However, in the case of active galactic
nuclei with a huge central black hole and an important magnetic field, the
Ernst spacetime can represent some relevant properties of the galactic
structures. Therefore, it could even be astrophysically important to discuss
and illustrate the properties of the Ernst spacetime.

The Ernst spacetime has the important property that it is not asymptotically
flat. Far from the black hole, the spacetime is closely related to Melvin's
magnetic universe, representing a cylindrically symmetric spacetime filled
with an uniform magnetic field only. The Ernst spacetime is axially symmetric,
and its structure corresponds to the structure of the Schwarzschild spacetime
only along its axis of symmetry. Off the axis, the differences are very
spectacular, and we shall demonstrate them for the equatorial plane, which is
the symmetry plane of the spacetime.

Motion of test particles and photons in the Ernst spacetime was discussed by
number of authors~\cite{DHV,E84,E85,AG,KV}, and number of properties different
from those of the Schwarzschild spacetime were discovered. For example, if the
magnetic field is weak enough, a stable circular photon orbit can exist behind
the unstable circular photon orbit located nearby the black-hole horizon. We
illustrate the impact of this fact on the structure of photon capture cones,
and we compare them with the corresponding ones in the Schwarzschild spacetime
in Section~\ref{cones}. Further we illustrate properties of the Ernst
spacetime by comparing the embedding diagrams of its $t={\rm const}$ sections with
embedding diagrams of the Schwarzschild spacetime both for ordinary geometry
and optical reference geometry in Section~\ref{embed}. These embedding
diagrams give in an illustrative way information on changes of the spacetime
structure caused by the presence of an uniform magnetic field about the
central black hole. The most important change comes from the asymptotic
behaviour of the spacetime which is manifested by the fact that the radius of
the embedding diagrams shrink asymptotically to zero. Finally, in
Section~\ref{concl}, we estimate which regions of the Ernst and Schwarzschild
spacetimes could be considered as having similar character.

\section{Photon capture cones}  \label{cones}

In the standard Schwarzschild coordinates $(t,r,\theta,\phi)$, and the
geometric system of units ($c=G=1$), the Ernst spacetime is determined by the
line element
\be
\fl \d s^2 = \Lambda^2 \left[-\left(1-\frac{2M}{r}\right)\d t^2 +
    \left(1-\frac{2M}{r}\right)^{-1}\d r^2 + r^2\,\d\theta^2 \right] +
    \frac{r^2 \sin^2 \theta}{\Lambda^2}\,\d\phi^2 \fcomma
\ee
where
\be
  \Lambda \equiv 1 + B^2 r^2 \sin^2 \theta \fcomma
\ee
with $M$ (mass) and $B$ (strength of the magnetic field) being the parameters
of the spacetime. The dimensionless combination of these parameters, $BM$, has
to be very small in astrophysically realistic situations.  Using, for a
moment, the c.g.s.\ units, we find
\be
  BM \sim 2\times 10^{-53} B_{\rm cgs}M_{\rm cgs} \fcomma
\ee
where $M = M_{\rm cgs}G/c^2$, and $B = B_{\rm cgs}G^{1/2}/c^2$. Then for a
black hole with $M = 10^9 M_{\odot}$, typical for active galactic nuclei, the
parameter $BM \sim 1$ could be obtained for an extremely strong magnetic field
of $B_{\rm cgs}\sim 10^{11}$\,Gauss.

In the following discussion we put $M=1$ for simplicity. Then the coordinates
$r,t$ become dimensionless, and $BM \rightarrow B$.

In the simple case of the motion of photons in the equatorial plane ($\theta =
\pi/2$), the geodesic equation $\D p^\mu/\d \lambda = 0$ with the
normalization condition $p^\mu p_\mu = 0$ can be solved easily. (The
4-momentum is $p^\mu=\d x^\mu/\d\lambda$, and $\lambda$ is the affine
parameter along the geodesic.) Except the constant latitude $\theta = \pi/2$,
there are constants of the motion connected to the time and azimuthal Killing
vectors, $\partial/\partial t$ and $\partial/\partial\phi$, of the spacetime:
\be
  -{\cal E} = p_t = g_{t\mu}p^\mu
\ee
and
\be
  \Phi = p_\phi = g_{\phi\mu}p^\mu.
\ee
Using the normalization condition, the equations of the photon motion can be
given in the following form
\bea
  (p^r)^2 = \left(\oder{r}{\lambda}\right)^2 =
    \frac{{\cal E}^2}{\Lambda^4} -
    \frac{1 - 2/r}{r^2} \Phi^2 \fcomma                     \label{radial}  \\
  p^\theta = \oder{\theta}{\lambda} = 0 \fcomma \\
  p^\phi = \oder{\phi}{\lambda} =
    \frac{\Lambda^2}{r^2}\,\Phi \fcomma \\
  p^t = \oder{t}{\lambda} =
    \frac{{\cal E}}{\Lambda^2 (1 - 2/r)}.
\eea

Introducing an impact parameter by the relation
\be
  l = \frac{\Phi}{{\cal E}} \fcomma
\ee
and making a rescaling of the affine parameter by ${\cal E}\lambda \rightarrow
\lambda$, we find from the radial equation (\ref{radial}) that the photon
motion is allowed in regions where
\be
  l^2 \leq l_{\rm R}^2(r;B) \equiv
    \frac{r^2}{\Lambda^4 \left(1-2/r \right)} .
\ee

The turning points of the radial motion of a photon with impact parameter $l$
are given by $l^2=l^2_{\rm R}(r;B)$.

As in the Schwarzschild spacetime, the function $l^2_{\rm R}(r;B)$ diverges at
$r=2$, giving location of the event horizon of the Ernst spacetime. However,
it behaves in a totally different manner asymptotically: $l_{\rm R}^2
\rightarrow 0$ for $r\rightarrow\infty$.

\begin{figure}[t]
\centering\leavevmode
\epsfxsize=.75\hsize \epsfbox{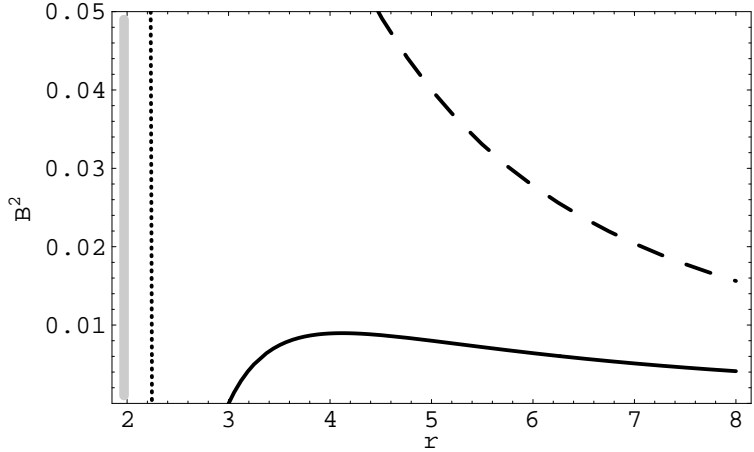}
\caption{The behaviour of  $B_{\rm ph}^2$ (solid curve), extreme of the
  ordinary embeddings $B_{\rm max}^2$ (dashed curve), defined by Eq.\
  \protect\ref{rmax}, and the radii of the limit of the embeddability of the
  optical geometry, determined by solving Eq.\ (\protect\ref{embedcond})
  numerically (dotted curve) as functions of $r$.}
\label{f1}
\end{figure}

Since 
\be
  \pder{l_{\rm R}^2}{r} =
    \frac{2r^2 (r - 3 + 5r^2B^2 - 3r^3B^2)}%
    {(1+B^2r^2)^5 (r-2)^2} \fcomma
\ee
the photon circular orbits, located at the local extrema of $l_{\rm R}^2$,
are determined by the condition
\be
  3B^2r^3 - 5B^2r^2 - r + 3 = 0 \fcomma
\ee
or, equivalently, by the relation
\be
  B^2 = B^2_{\rm ph}(r) \equiv \frac{r-3}{r^2(3r-5)}.
\ee
The function $B^2_{\rm ph}(r)$ is given in Fig.\ \ref{f1}. It is defined at
$r\geq 3$; as expected, for $B=0$ we obtain the circular photon orbit of the
Schwarzschild spacetime. Its maximum is located at $r_{\rm c} =
(8+\sqrt{19})/3$, and it defines a critical value of the magnetic-field
parameter:
\be
  B^2_{\rm c} =
    \frac{3(\sqrt{19} - 1)}{(8 + \sqrt{19})^2(3 + \sqrt{19})}
    \approx 0.008965 \fcomma
\ee
or
\be
  B_{\rm c} \approx 0.0947.
\ee

\begin{figure}[t]
\centering\leavevmode
\epsfxsize=.75\hsize \epsfbox{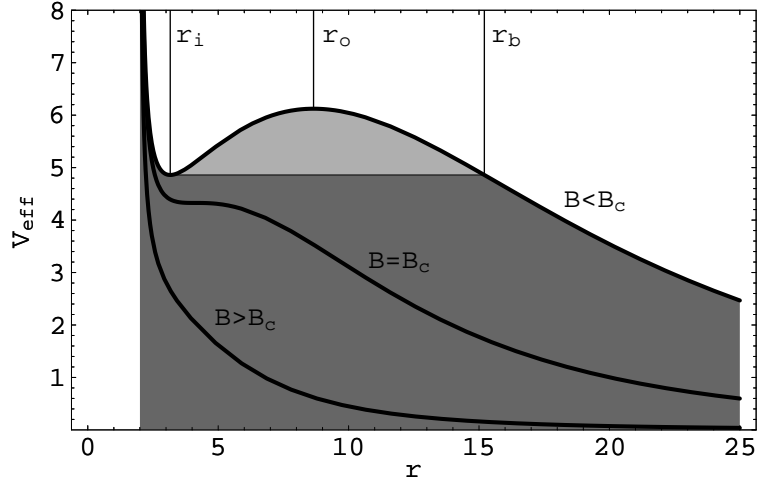}
\caption{The function $l_{\rm R}^2(r;B)$ governing the photon motion is
  drawn for $B=0.2>B_{\rm c}$, $B=B_{\rm c}\approx 0.0947$ and $B=0.06<B_{\rm
  c}$. Photons having its radius of emission $r$ and the impact parameter $l$
  in the dark grey area are captured, the light grey area represents bound
  photons.}
\label{f2}
\end{figure}

If $B < B_{\rm c}$, there exist two circular photon orbits. The inner one at
$r_{\rm i}$ is unstable with respect to radial perturbations, the outer one at
$r_{\rm o}$ is stable. If $B = B_{\rm c}$, these two circular orbits coalesce.
If $B > B_{\rm c}$, no photon circular orbits can exist (see Fig.\ \ref{f2}).
For realistic magnetic fields, $B \ll 1$, we find
\be
  r_{\rm i} \sim 3,\;  r_{\rm o} \sim \frac{2}{\sqrt{6}\,B}.
\ee 
We can see that if $B < B_{\rm c}$, bound photons orbits exist around the
outer, stable, photon circular orbit, which are not captured by the hole, and
cannot escape to infinity. Such orbits can extend down to the unstable photon
circular orbit. On the other hand, purely radially directed photons with a
zero impact parameter can only escape to infinity, if they are outward
directed initially. If $B > B_{\rm c}$, all photons are captured by the hole,
except the outward directed radial photons, which escape to infinity.

\begin{figure}[t]
\centering
\begin{minipage}{.24\hsize}
\centering\leavevmode\epsfxsize=\hsize \epsfbox{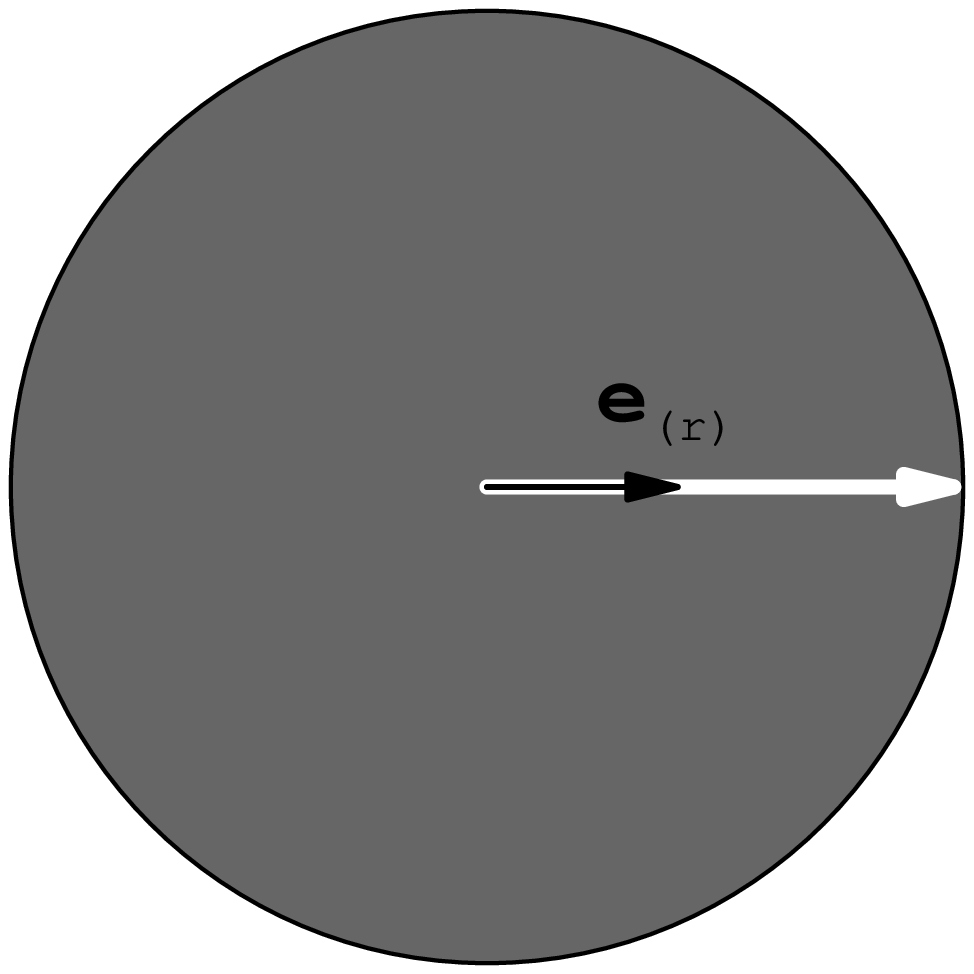}
\par\vskip 1mm\par $r\leq r_{\rm i}\approx 3.161$
\end{minipage}\hfill%
\begin{minipage}{.24\hsize}
\centering\leavevmode\epsfxsize=\hsize \epsfbox{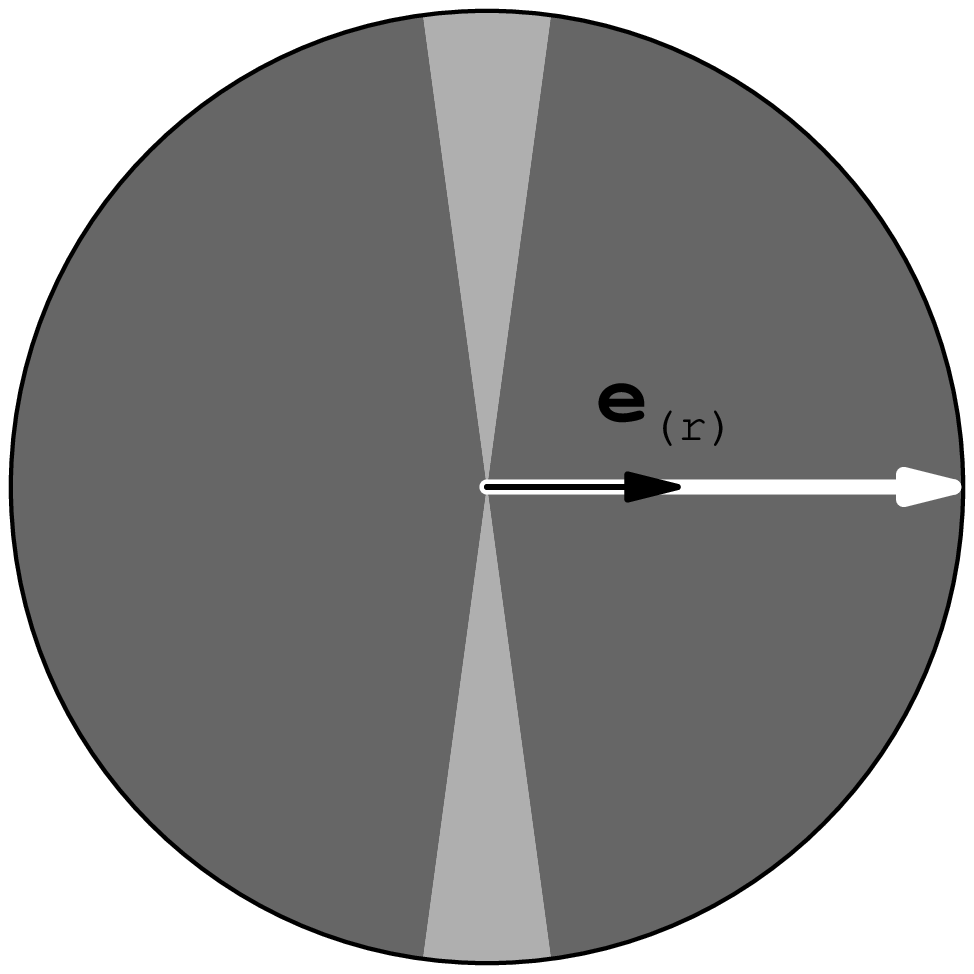}
\par\vskip 1mm\par $r=3.5$
\end{minipage}\hfill%
\begin{minipage}{.24\hsize}
\centering\leavevmode\epsfxsize=\hsize \epsfbox{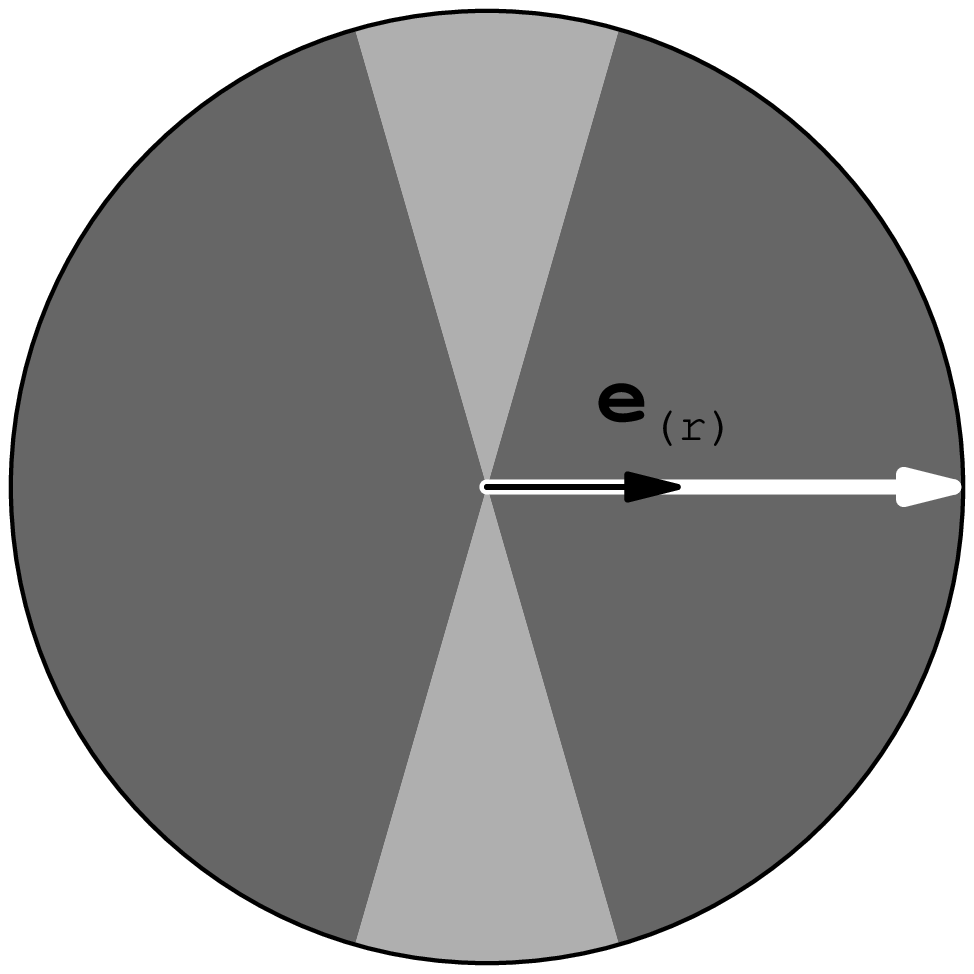}
\par\vskip 1mm\par $r=4$
\end{minipage}\hfill%
\begin{minipage}{.24\hsize}
\centering\leavevmode\epsfxsize=\hsize \epsfbox{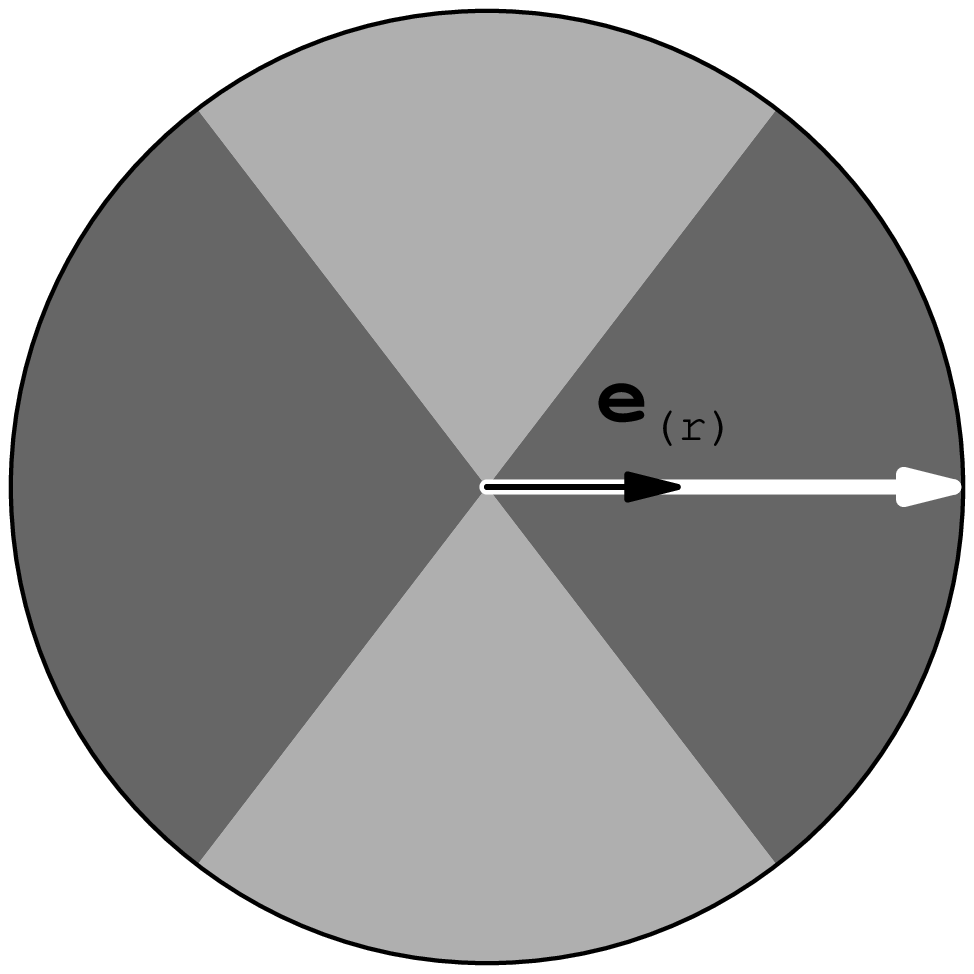}
\par\vskip 1mm\par $r=r_{\rm o}\approx 8.656$
\end{minipage}
\par\vskip 5mm\par
\begin{minipage}{.24\hsize}
\centering\leavevmode\epsfxsize=\hsize \epsfbox{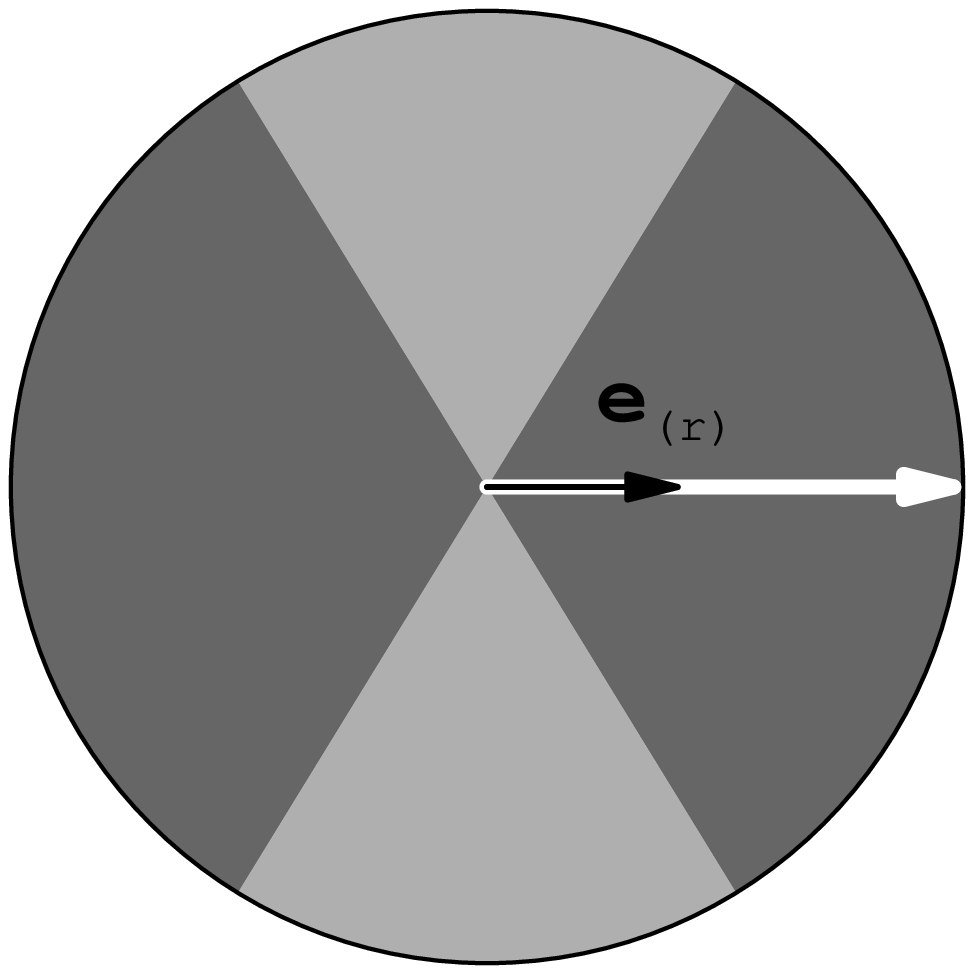}
\par\vskip 1mm\par $r=12$
\end{minipage}\hfill%
\begin{minipage}{.24\hsize}
\centering\leavevmode\epsfxsize=\hsize \epsfbox{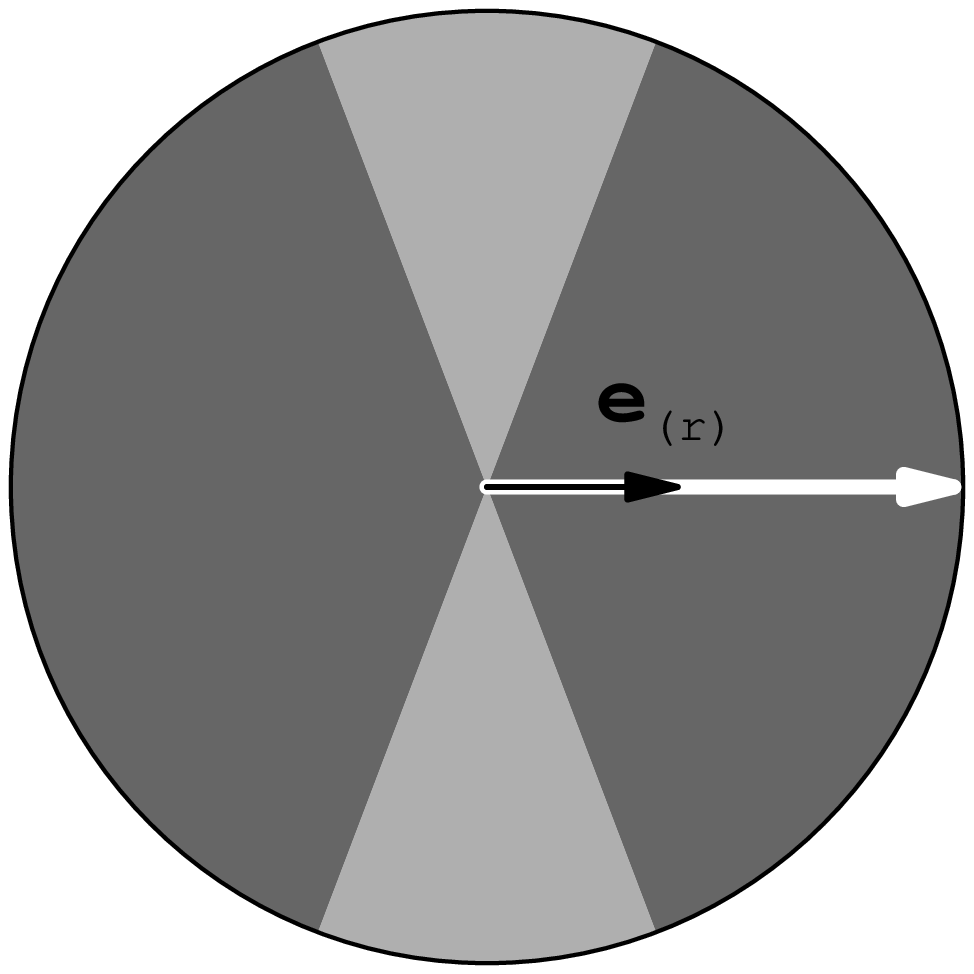}
\par\vskip 1mm\par $r=14$
\end{minipage}\hfill%
\begin{minipage}{.24\hsize}
\centering\leavevmode\epsfxsize=\hsize \epsfbox{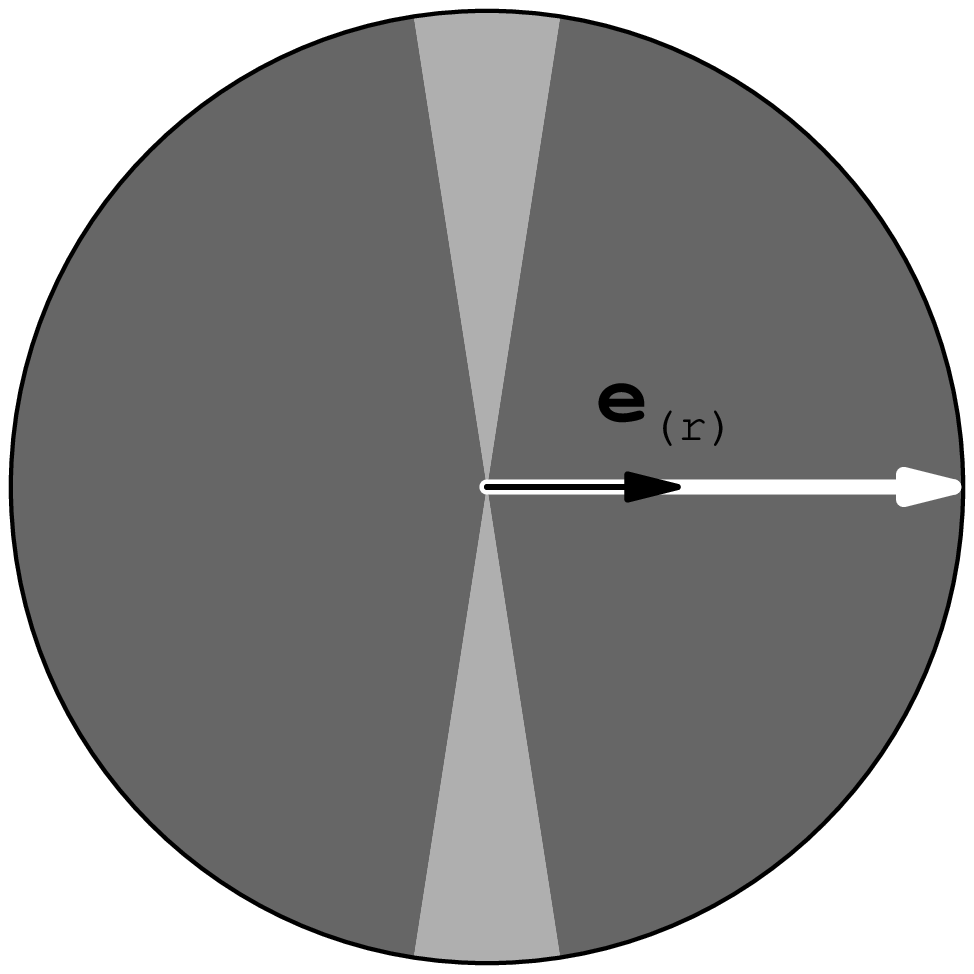}
\par\vskip 1mm\par $r=15$
\end{minipage}\hfill%
\begin{minipage}{.24\hsize}
\centering\leavevmode\epsfxsize=\hsize \epsfbox{f3ah.eps}
\par\vskip 1mm\par $r\geq r_{\rm b}\approx 15.205$
\end{minipage}
\caption{Directions of captured (dark grey) and bound (light grey) photons for
  several values of the radius of a static observer for $B=0.06<B_{\rm c}$.
  Note that if $2<r\leq r_{\rm i}\approx 3.161$ or $r\geq r_{\rm b}\approx
  15.205$, all photons are captured. Only the purely radially outward emitted
  photons (white arrow) can escape to infinity. Escape angles for the
  Schwarzschild black hole can be found, e.g., in~\protect\cite{MTW}.}
\label{f3}
\end{figure}

Now, we can determine the photon capture cones of static observers. In the
Ernst spacetime, we introduce the tetrad of differential forms
$\omega^{(\alpha)}_\mu$ related to the static observers by the relations
\bea
  \omega^{(t)} = (1+B^2r^2)\sqrt{1- \frac{2}{r}}\,\,\d t \fcomma \\
  \omega^{(r)} = (1+B^2r^2)\left(1-\frac{2}{r}\right)^{-1/2}\d r \fcomma \\
  \omega^{(\theta)} = (1+B^2r^2) r\, \d\theta \fcomma \\
  \omega^{(\phi)} = \frac{r}{1+B^2r^2}\,\d\phi.
\eea 
The tetrad of vectors $e^\mu_{(\alpha)}$, dual to the tetrad of the
differential forms, is related in the standard manner and will not be
explicitly given here. The components of photon's 4-momentum, as measured by
the static observers, are determined by the relations
\bea
  p_{(\alpha)} = p_\mu e^\mu_{(\alpha)} \fcomma \\
  p^{(\alpha)} = p^\mu \omega^{(\alpha)}_\mu.
\eea
and its directional angle $\psi$, related to the outward radial direction, is
given by the relation
\be
  \sin\psi = \frac{p^{(\phi)}}{p^{(t)}}.
\ee
Since the relevant components of the 4-momentum are
\bea
  p^{(t)} = \frac{{\cal E}}{(1+B^2r^2)\sqrt{1-2/r}} \fcomma \\
  p^{(\phi)} = \frac{(1+B^2 r^2)}{r}\,\Phi \fcomma
\eea
we arrive at the final formula
\be
  \sin\psi = (1+B^2r^2)\sqrt{1-\frac{2}{r}}\; \frac{l}{r}.
\ee

The photon capture cones can be determined by using the function $l^2_{\rm
R}(r;B)$ governing the photon motion. In establishing the directional angle of
marginally captured photons $\psi_{\rm c}$ for $B < B_{\rm c}$, we have to use
impact parameter $l_{\rm c}$ of the unstable circular photon orbit, which
plays a crucial role. Directions of captured and bound photons are determined
by a numerical procedure, and they are depicted in a sequence of figures for
typical values of the radius of static observers in the field of a black hole
with the parameter $B=0.06<B_{\rm c}$ (see Fig.\ \ref{f3}). Notice that for
impact parameters satisfying the condition $0<l^2 <l^2_{\rm c}$, all the
photons are captured by the hole. The bound orbits can exist between the radii
$r_{\rm i}$ and $r_{\rm b}$; the definition of the radius $r_{\rm b}$ is clear
from Fig.\ \ref{f2}. At radii $r_{\rm i} < r < r_{\rm b}$, regions
corresponding to bound orbits appear around directions orthogonal to the
outward radial direction. The regions are most extended at $r = r_{\rm o}$,
and they shrink to zero for $r \rightarrow r_{\rm i}$ and $r \rightarrow
r_{\rm b}$.

If $B > B_{\rm c}$, the situation is quite simple. At each radii there are
only captured photons, except the purely radially outward directed photons,
which reach infinity asymptotically.

\section{Embedding diagrams}  \label{embed}

It is well known that embedding diagrams of some representative sections of
black-hole spacetimes give an useful intuitive understanding of the basic
properties of these spacetimes (see, e.g., \cite{MTW,C,KSA,SH}). Because we
are familiar to the Euclidean space, usually two-dimensional sections of the
black-hole spacetimes are embedded into the three-dimensional space with the
line element expressed in the cylindrical coordinates:
\be
  \d s^2 = \d\rho^2 + \rho^2\,\d\phi^2 + \d z^2.
\ee 
Of course, embeddings into some other spaces can also give interesting
information. However, here we restrict our attention to the most
straightforward Euclidean case.

\begin{figure}[p]
\def\ewdth{0.35}
\def\fwdth{0.2}
\centering
\begin{minipage}{.94\hsize}
\centering
\begin{minipage}{.855\hsize}
\begin{minipage}{\ewdth\hsize}
\centering\leavevmode\epsfxsize=\hsize \epsfbox{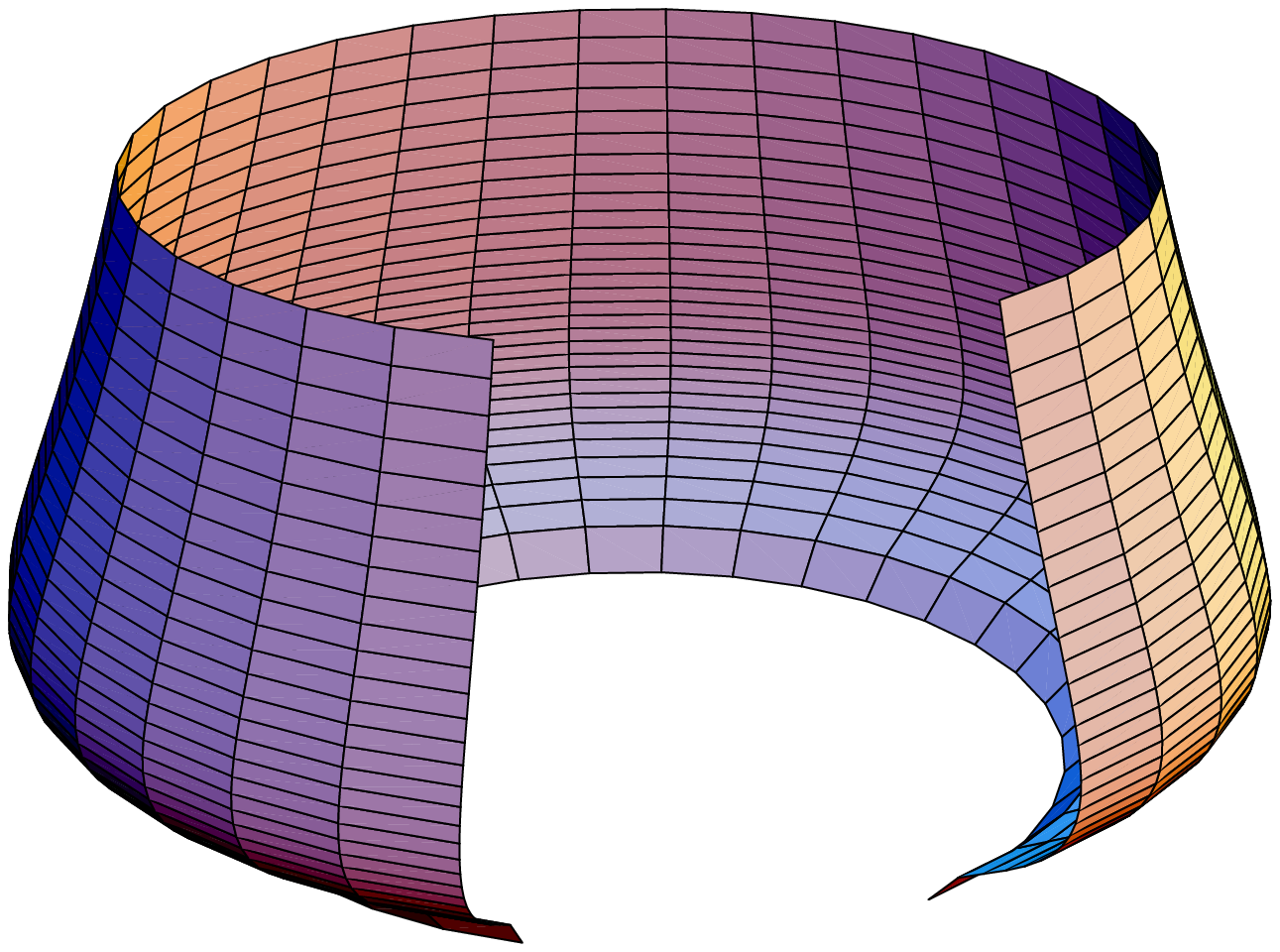}
\end{minipage}\hfill%
\begin{minipage}{\fwdth\hsize}
\centering $B=0.2$
\end{minipage}\hfill%
\begin{minipage}{\ewdth\hsize}
\centering\leavevmode\epsfxsize=\hsize \epsfbox{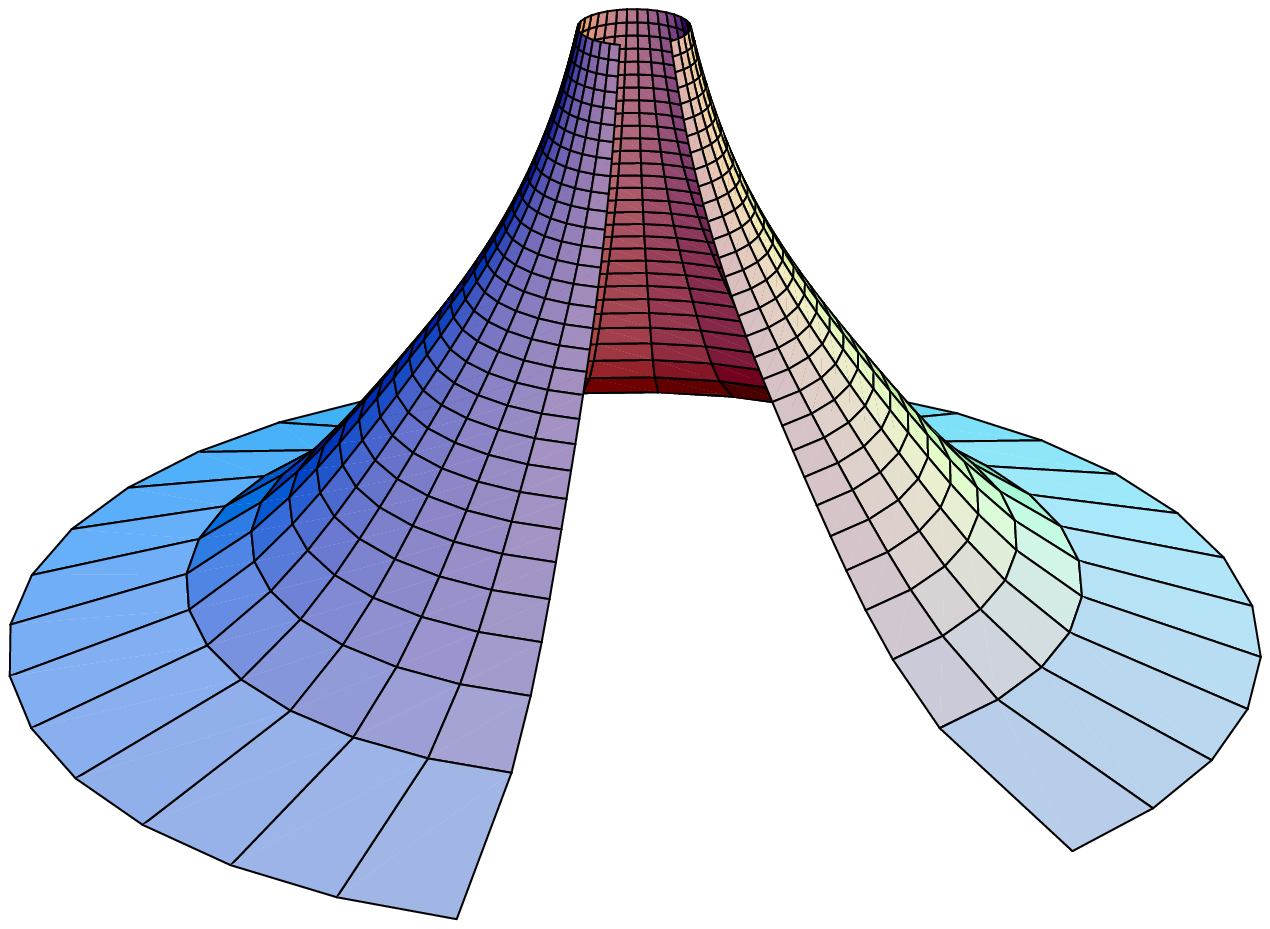}
\end{minipage}
\par\vskip 5pt\par
\begin{minipage}{\ewdth\hsize}
\centering\leavevmode\epsfxsize=\hsize \epsfbox{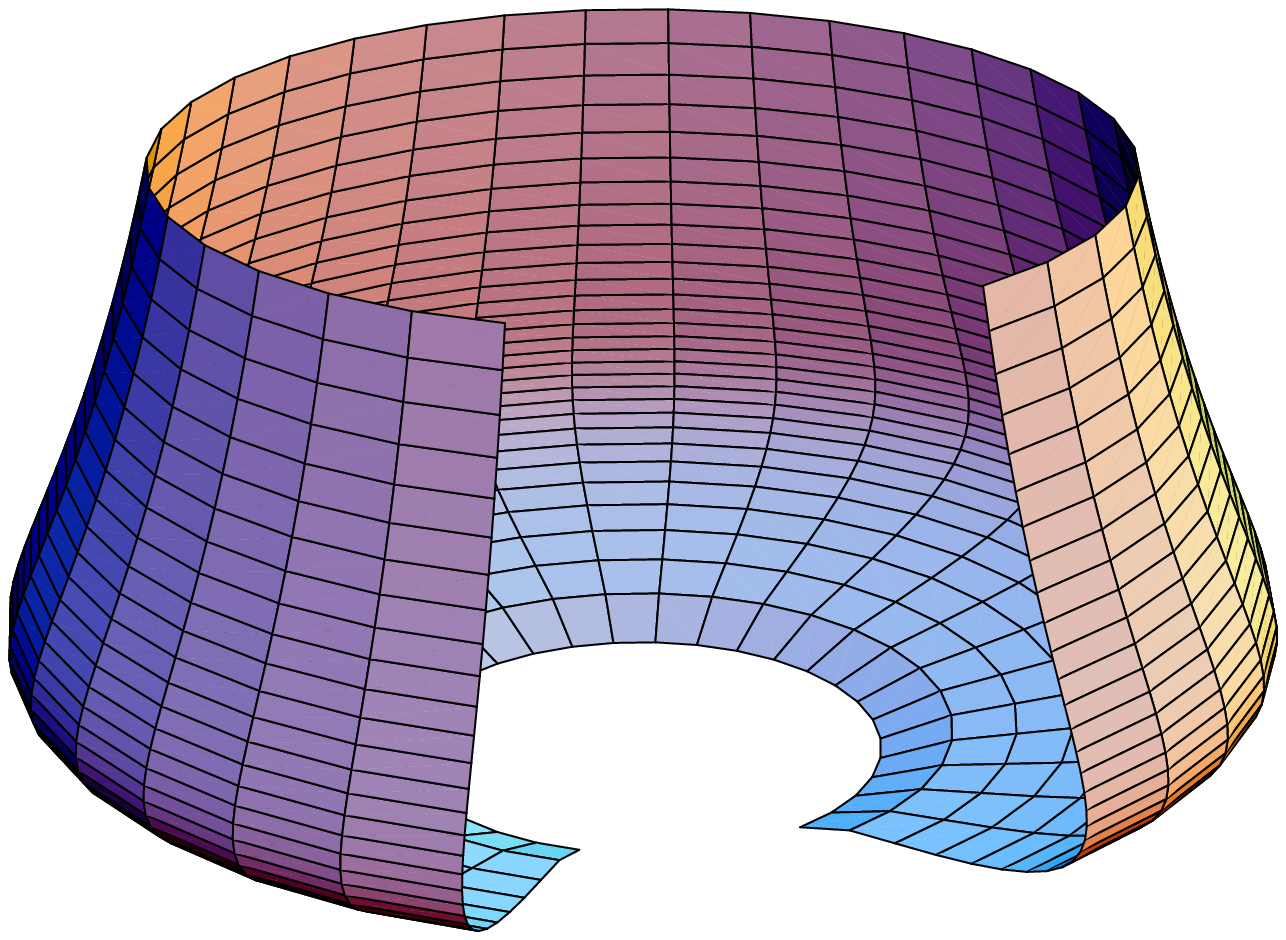}
\end{minipage}\hfill%
\begin{minipage}{\fwdth\hsize}
\centering $B=0.1$
\end{minipage}\hfill%
\begin{minipage}{\ewdth\hsize}
\centering\leavevmode\epsfxsize=\hsize \epsfbox{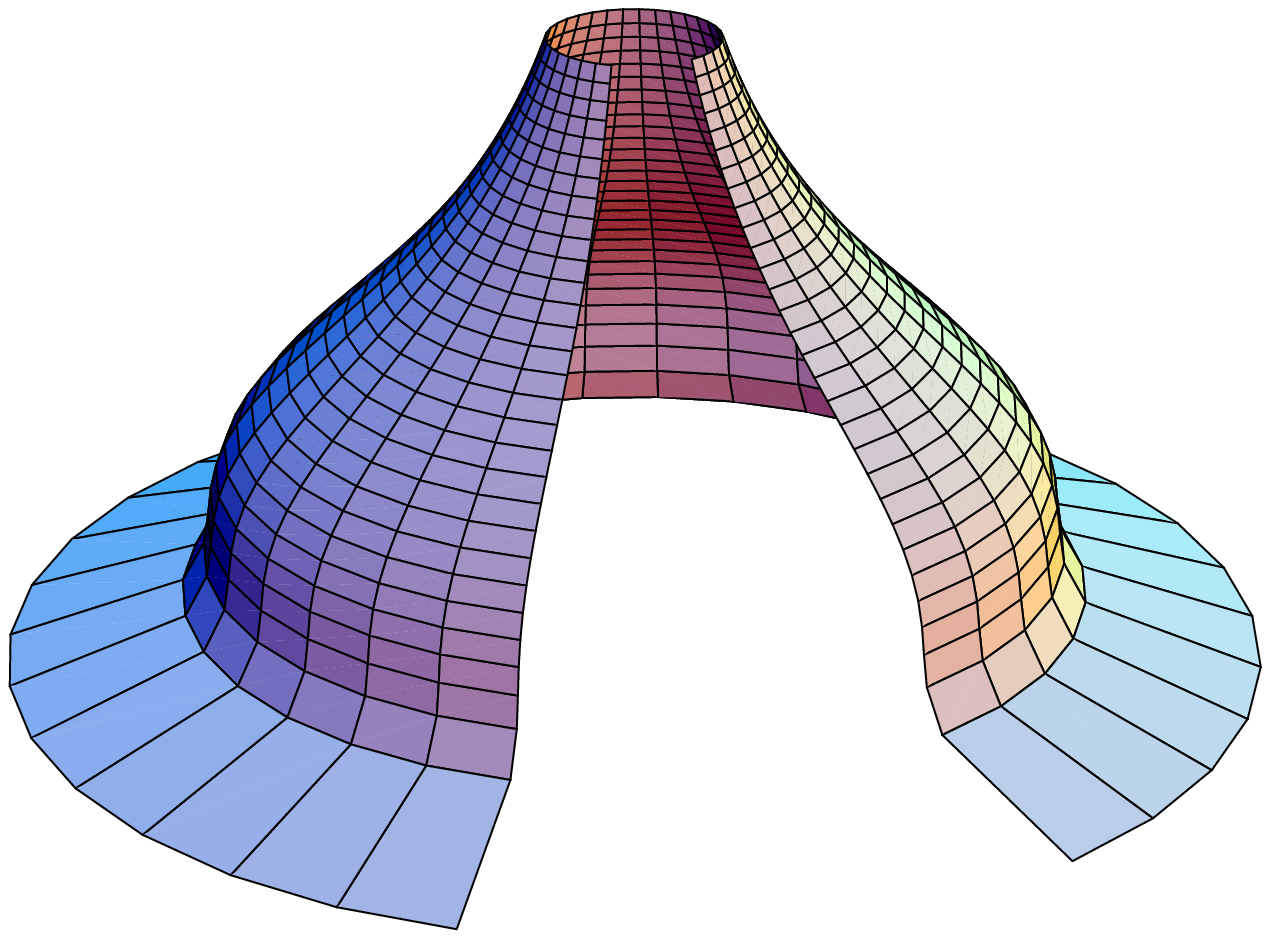}
\end{minipage}
\par\vskip 5pt\par
\begin{minipage}{\ewdth\hsize}
\centering\leavevmode\epsfxsize=\hsize \epsfbox{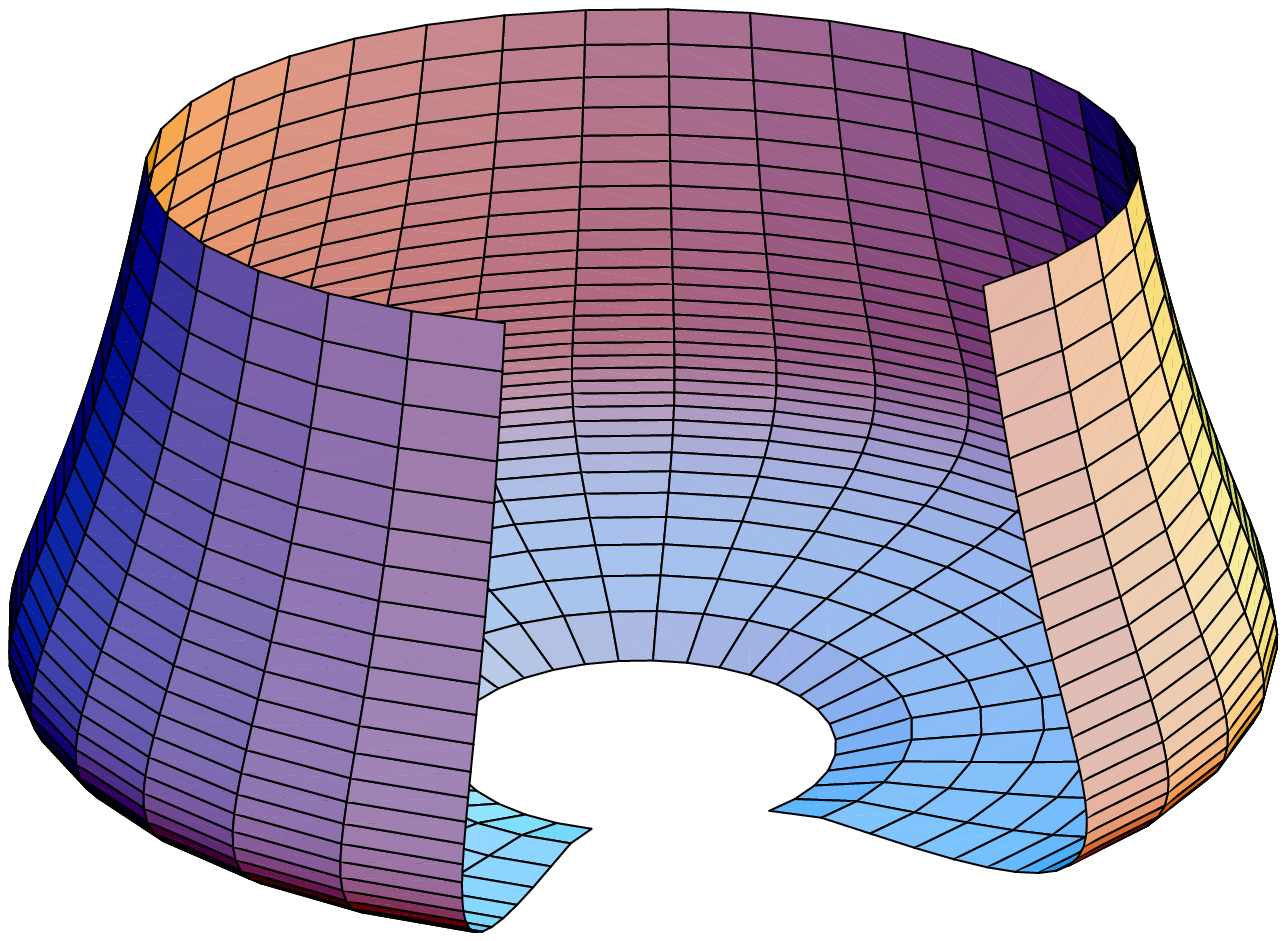}
\end{minipage}\hfill%
\begin{minipage}{\fwdth\hsize}
\centering $B=0.08$
\end{minipage}\hfill%
\begin{minipage}{\ewdth\hsize}
\centering\leavevmode\epsfxsize=\hsize \epsfbox{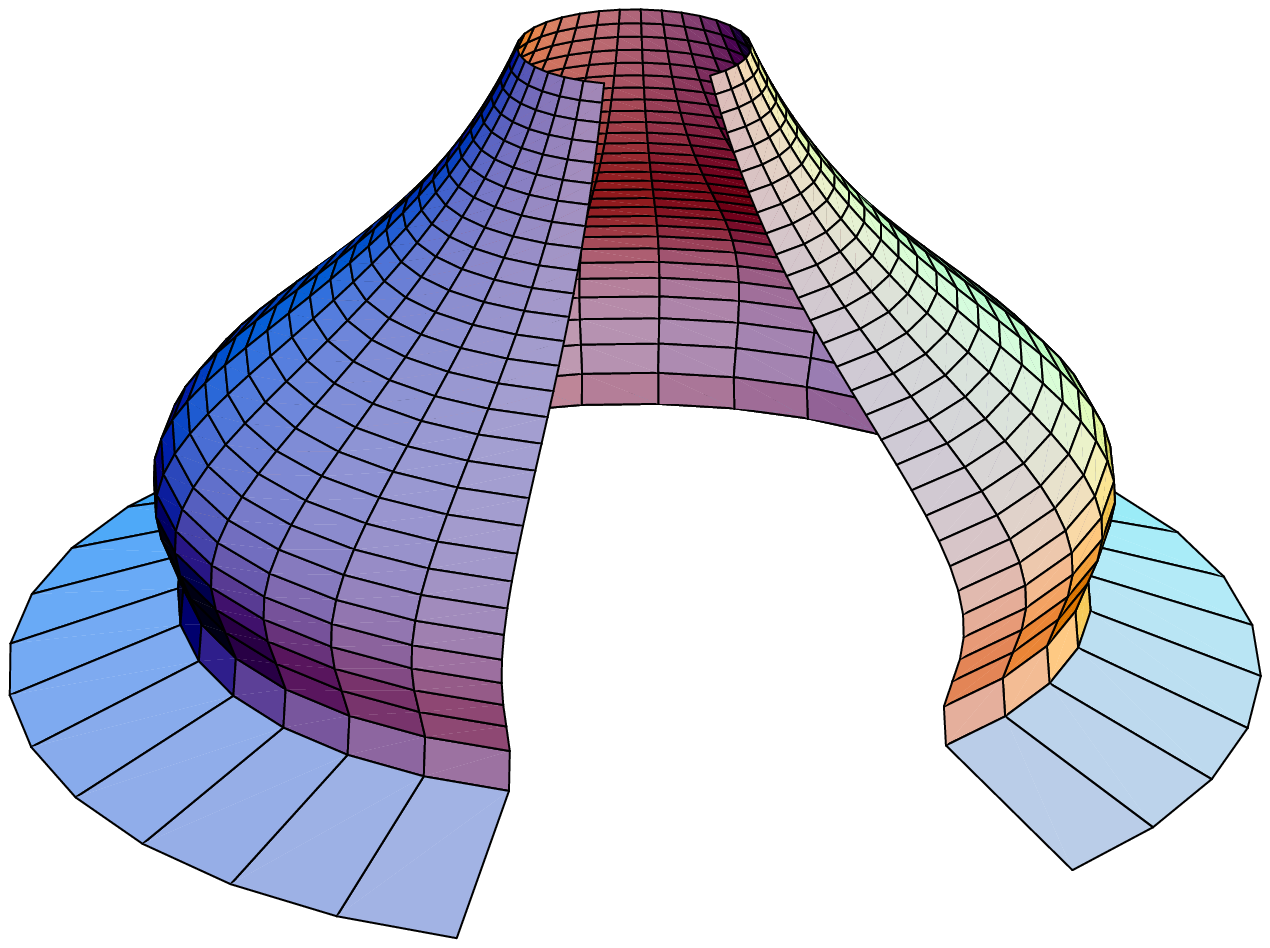}
\end{minipage}
\par\vskip 5pt\par
\begin{minipage}{\ewdth\hsize}
\centering\leavevmode\epsfxsize=\hsize \epsfbox{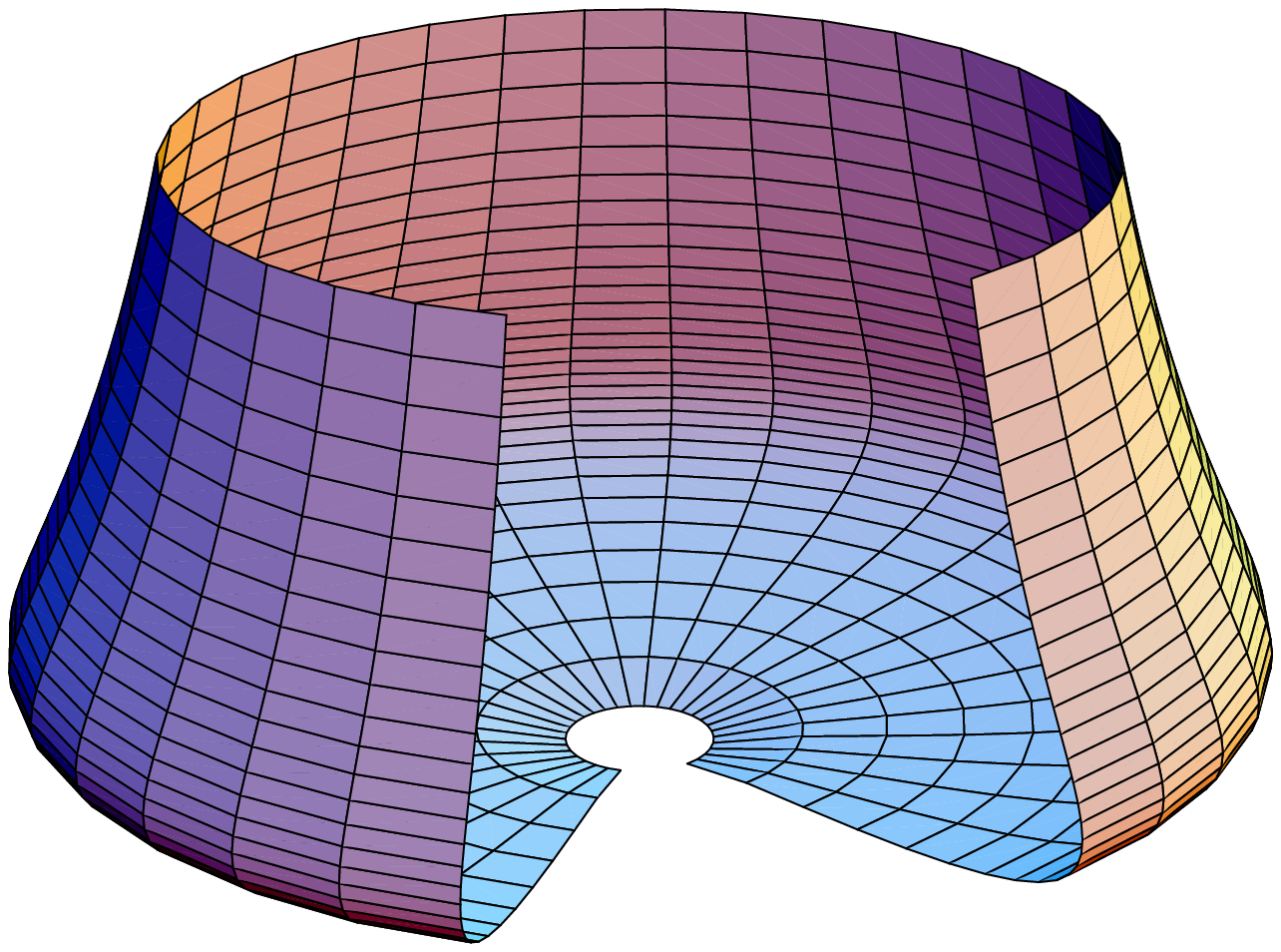}
\end{minipage}\hfill%
\begin{minipage}{\fwdth\hsize}
\centering $B=0.03$
\end{minipage}\hfill%
\begin{minipage}{\ewdth\hsize}
\centering\leavevmode\epsfxsize=\hsize \epsfbox{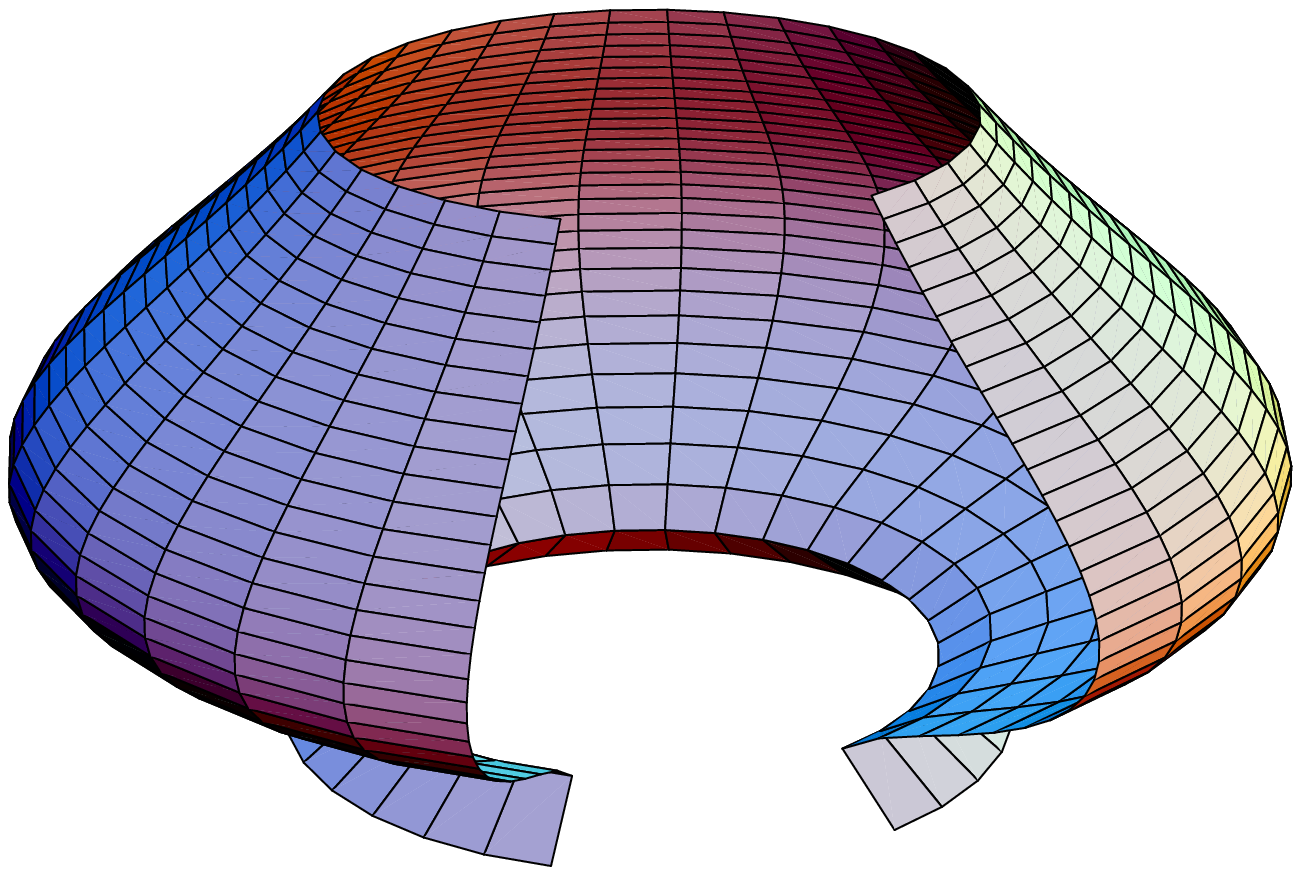}
\end{minipage}
\par\vskip 5pt\par
\begin{minipage}{\ewdth\hsize}
\centering\leavevmode\epsfxsize=\hsize \epsfbox{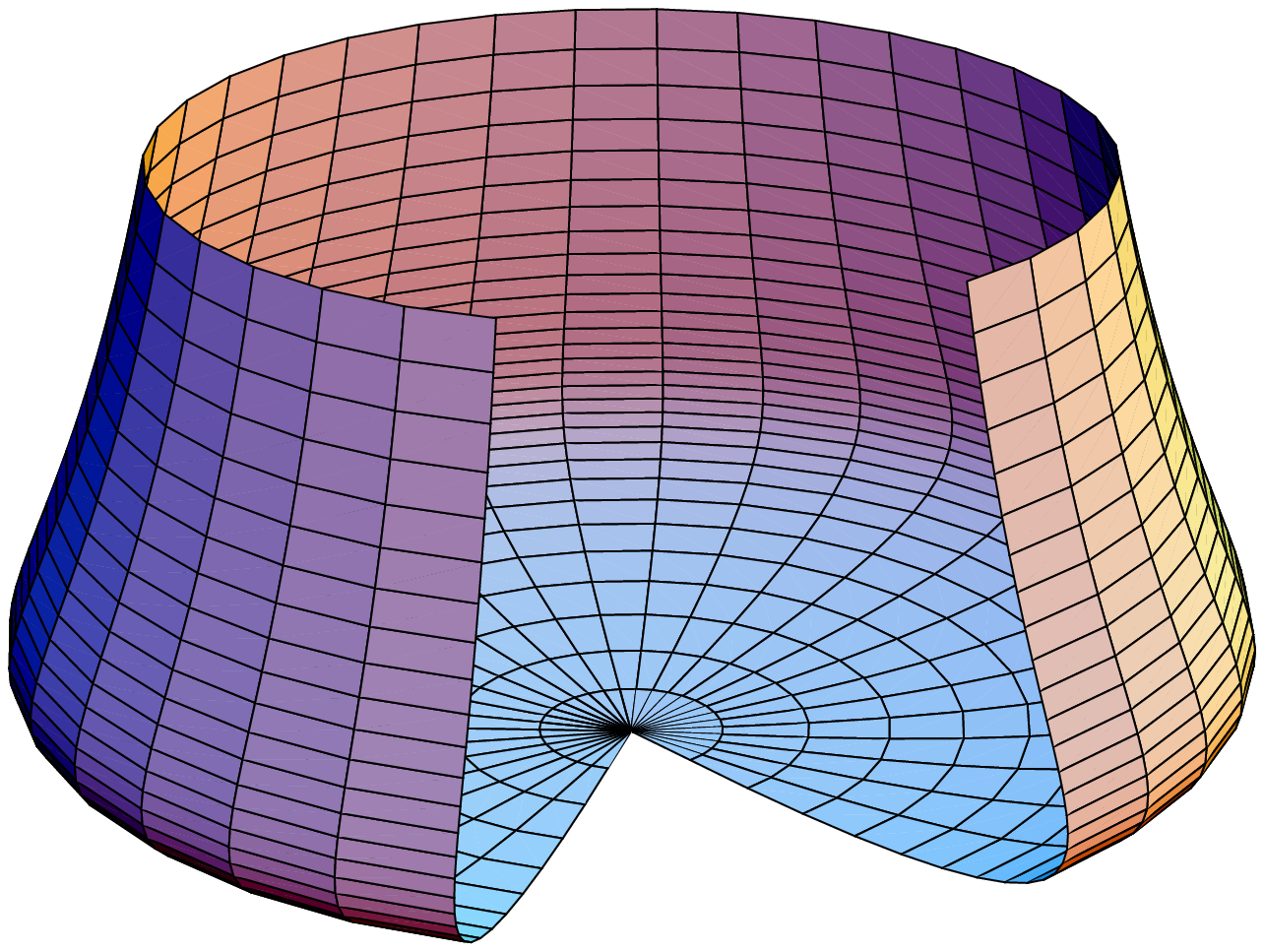}
\end{minipage}\hfill%
\begin{minipage}{\fwdth\hsize}
\centering $B=10^{-3}$
\end{minipage}\hfill%
\begin{minipage}{\ewdth\hsize}
\centering\leavevmode\epsfxsize=\hsize \epsfbox{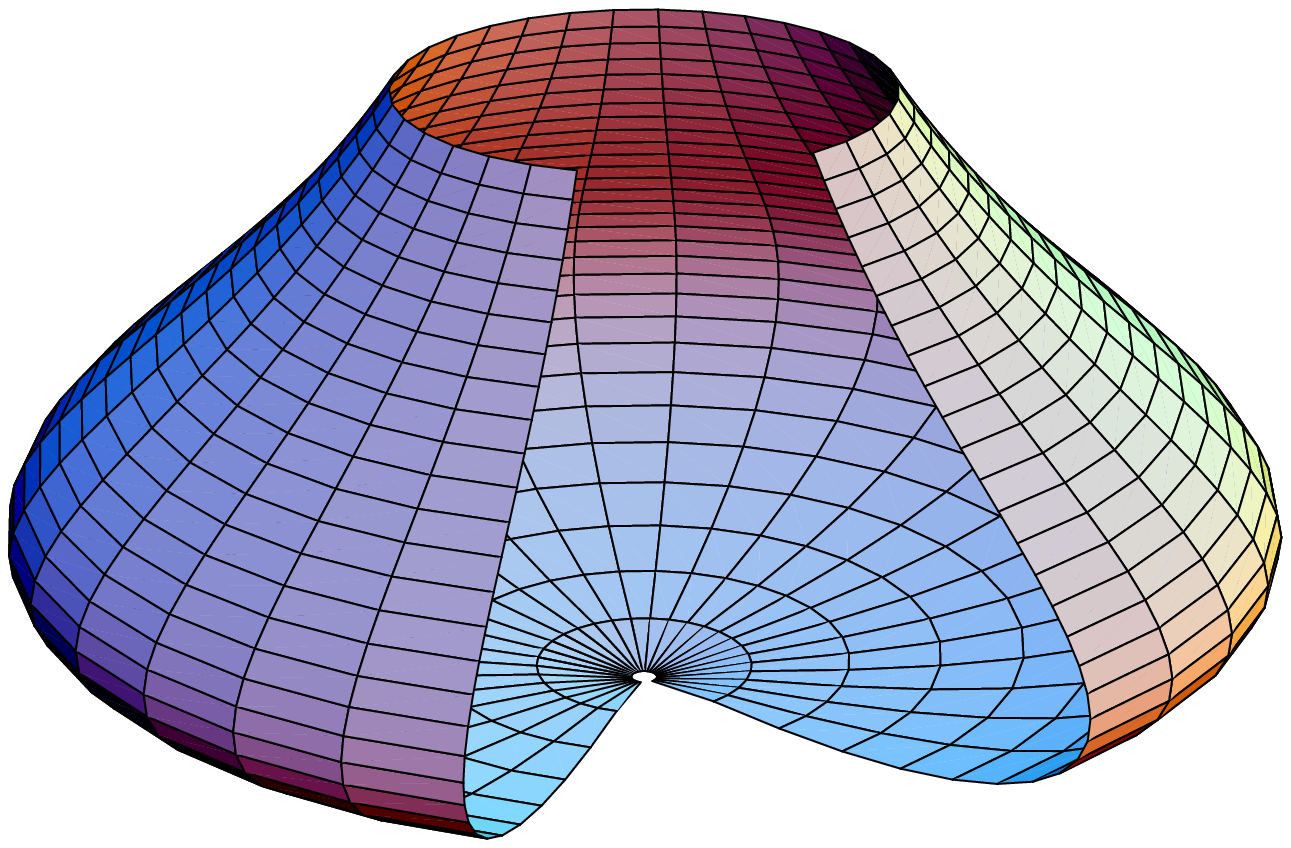}
\end{minipage}
\end{minipage}
\end{minipage}
\caption{Series of embedding diagrams of the ordinary space (left column) and
  the optical space (right column). From the optical embeddings one can easily
  see that for $B>B_{\rm c}$ (first two top rows) no circular photon orbits
  exist, for $B<B_{\rm c}$ (third and fourth row from top) and $B\ll B_{\rm
  c}$ (bottom row) stable and unstable circular photon orbit exist. In the
  last case, the opening and the `fold' in the lower part of the `cup',
  representing the unstable circular orbit are extremally small and beyond the
  image resolution. Note that both $\rho$ and $z$ scales are adjusted so that
  all the plots occupy the same area. For real proportions, see Fig.\
  \protect\ref{f5} (ordinary embeddings) and Fig.\ \protect\ref{f6} (optical
  embeddings).}
\label{f4}
\end{figure}

\begin{figure}[t]
\centering\leavevmode
\epsfxsize=.7\hsize \epsfbox{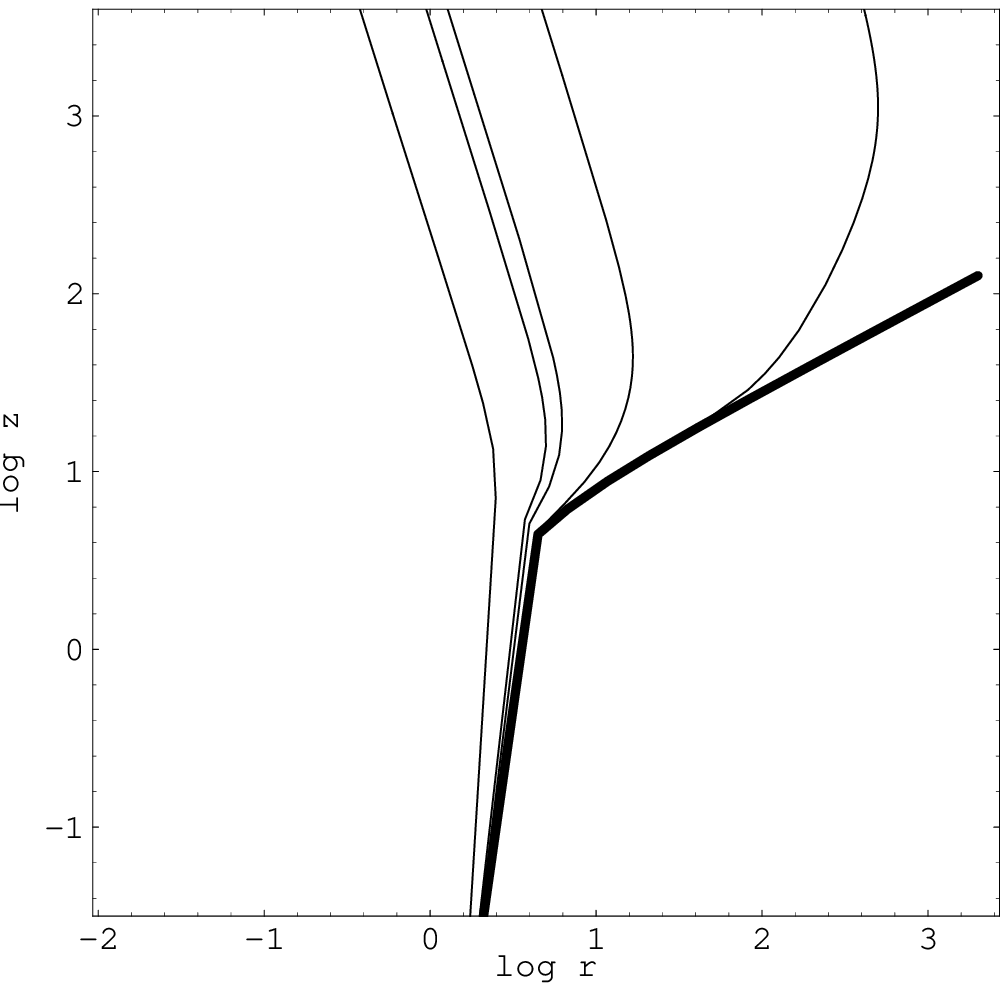}
\caption{The behaviour of the embedding function $z(\rho)$ for Ernst ordinary
  space (axial cuts of the diagrams in the left column of Fig.\
  \protect\ref{f4}, thin curves) in comparison with the pure Schwarzschild
  case (bold curve), in log-log coordinates that enable to show the order
  differences in size. The Ernst embedding functions are given (left to right)
  for the following values of the parameter $B$: 0.2, 0.1, 0.08, 0.03, and
  $10^{-3}$.}
\label{f5}
\end{figure}

The time Killing vector field $\partial/\partial t$ of the Ernst spacetime
leads to a privileged space-like sections---namely the hypersurfaces $t = {\rm
const}$. We shall do the embedding for the equatorial plane of these
hypersurfaces. Then the line element reads
\be                                                            \label{eqplane}
  \d l^2  =
    \frac{(1+B^2r^2)^2}{1-2/r}\, \d r^2 +
    \frac{r^2}{(1+B^2r^2)^2}\, \d\phi^2.
\ee
We have to find a surface $z = z(\rho)$ in the three dimensional Euclidean
space, which is isometric to the two-dimensional space given by
(\ref{eqplane}). Therefore, we identify the space with the line element
\be                                                               \label{euclid}
  \d l^2_{({\rm E})} =
    \left[1 + \left(\oder{z}{\rho}\right)^2 \right]\d\rho^2 +
    \rho^2\,\d\phi^2 .
\ee
Clearly, the azimuthal coordinates can be identified. But, contrary to the
Schwarz\-schild case, the radial coordinates cannot be identified directly.
There is
\be
  \rho = \frac{r}{1+B^2r^2} \fcomma
\ee
and due to this relation the Ernst and Schwarzschild embeddings are totally
different in the asymptotic region of $r \rightarrow \infty$. Really, $\rho
\rightarrow 0$ for $r \rightarrow \infty$, and the Ernst embedding diagram
shrinks to zero radius asymptotically. Because there is
\be                                                              \label{drhodr}
  \oder{\rho}{r} = \frac{1-B^2r^2}{(1+B^2r^2)^2} \fcomma
\ee
the maximum radius of the embedding diagram corresponds to an Ernst radial
coordinate
\be                                                              \label{rmax}
  r_{\rm max}(B) = \frac{1}{B} \fcomma
\ee
and an Euclidean radial coordinate
\be                                                              \label{rhomax}
  \rho_{\rm max}(B) = \frac{1}{2B}.
\ee
The embedding formula $z = z(\rho)$ is given by the relation
\be                                                              \label{dzdrho}
  \left(\oder{z}{\rho}\right)^2 =
    \frac{(1+B^2r^2)^2}{1- 2/r}
    \left(\oder{r}{\rho}\right)^2 - 1.
\ee
However, it is much simpler to express the embedding formula in a parametric
form, using $r$ as a parameter. From (\ref{drhodr}) and (\ref{dzdrho}) we
arrive at
\be
  \left(\oder{z}{r} \right)^2 =
    \frac{(1+B^2r^2)^6 - (1-B^2r^2)^2 \left(1- 2/r \right)}%
    {\left(1-2/r\right) (1+B^2r^2)^4}.
\ee
The embedding (or reality) condition $(\d z/\d r)^2 \geq 0$ implies two
relations to be satisfied:
\bea
  (1+B^2r^2)^6 - (1-B^2r^2)^2 \left(1-\frac{2}{r}\right) \geq 0 \fcomma \\
  r - 2 \geq 0.
\eea
A numerical analysis shows that the first condition is satisfied at $r \geq
2$ for arbitrary $B^2 > 0$. The embedding diagrams has the same character for
all values of the parameter $B$. We present them in Fig.\ \ref{f4}. For
comparison with the Schwarzschild case, we give the functions $z(\rho)$ in
Fig.\ \ref{f5}.

\begin{figure}[p]
\centering\leavevmode
\epsfxsize=.7\hsize \epsfbox{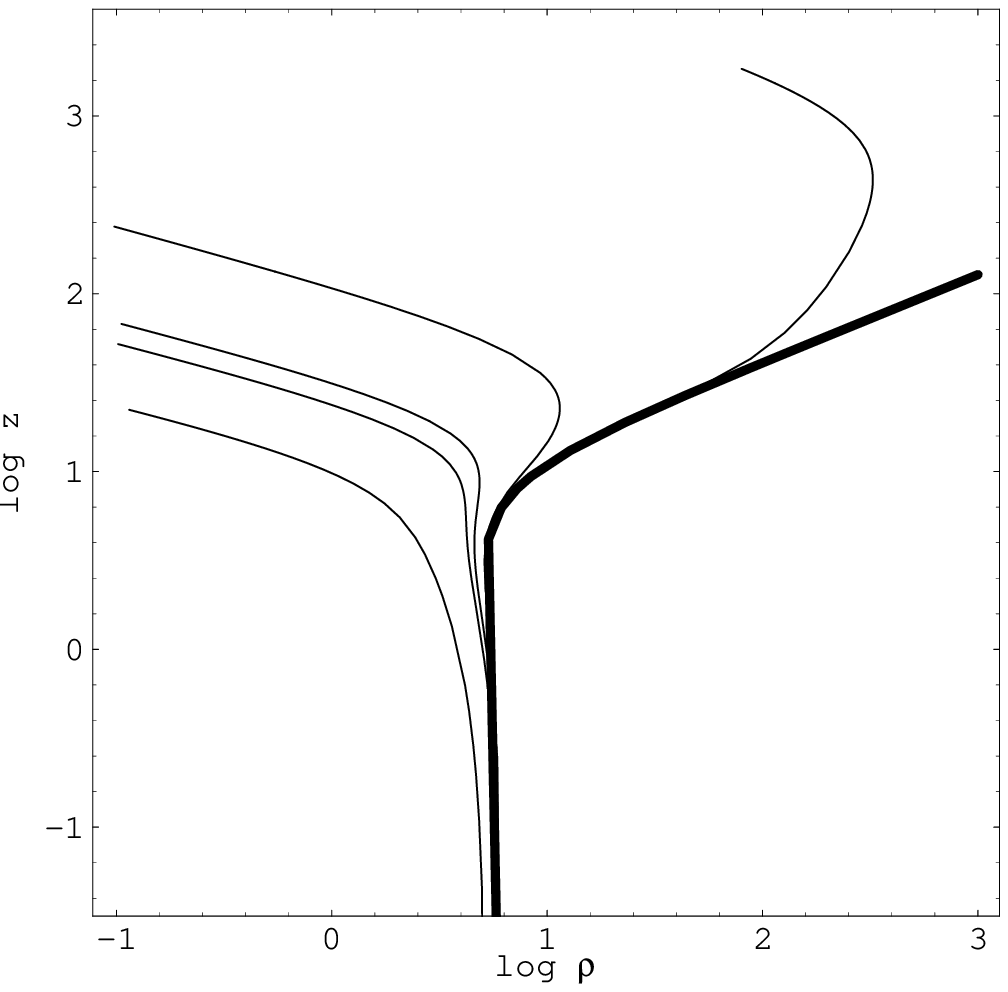}
\caption{The behaviour of the embedding function $z(\rho)$ for the optical
  space (axial cuts of the diagrams in the right  column of Fig.\
  \protect\ref{f4}, thin curves) in comparison with the pure Schwarzschild
  case (bold curve), in log-log coordinates that enable us to show the order
  differences in size. The Ernst embedding functions are given (left to right)
  for the following values of the parameter $B$: 0.2, 0.1, 0.08, 0.03, and
  $10^{-3}$.}
\label{f6}
\end{figure}

\begin{figure}[p]
\centering\leavevmode
\epsfxsize=.85\hsize \epsfbox{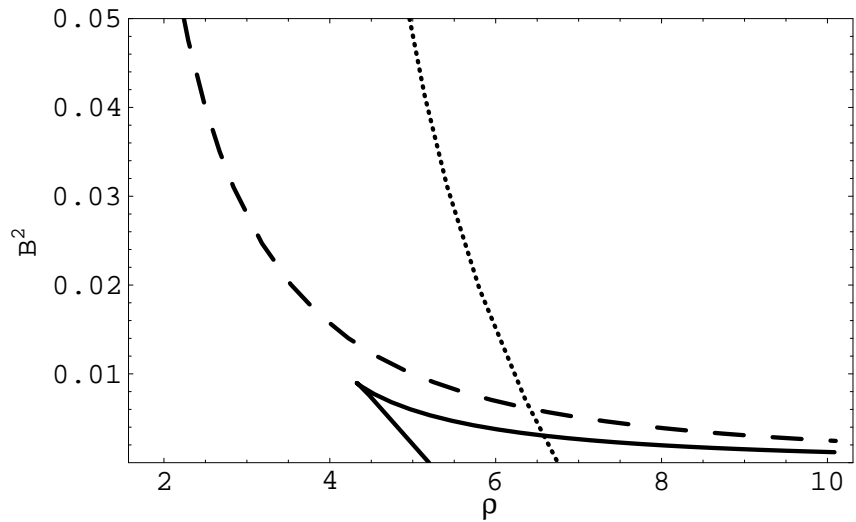}
\caption{Same as in Fig.\ \protect\ref{f1} in terms of the Euclidean
  coordinate $\rho$.}
\label{f7}
\end{figure}

Many relevant properties of black-hole spacetimes are related to the optical
reference geometry \cite{ACL,AP,A90,A92,S,AMS}. The optical geometry can be
introduced by the (3+1) conformal splitting
\be
  \d s^2 = \expon{2\Psi}(-\d t^2 + \d\tilde{l}^2) \fcomma
\ee
where 
\be
  \d\tilde{l}^2 = \tilde{g}_{ik}\,\d x^i \d x^k \fcomma
\ee
and $i,k = 1,2,3$. In the case of the Ernst metric we find
\bea
  \expon{2\Psi} = \Lambda^2 \left(1 - \frac{2}{r}\right) \fcomma \\
  \tilde{g}_{rr} = \left(1 - \frac{2}{r}\right)^{-2} \fcomma  \\
  \tilde{g}_{\theta\theta} = r^2 \left(1 - \frac{2}{r}\right) \fcomma  \\
  \tilde{g}_{\phi\phi} =
    \frac{r^2 \sin^2\theta}{\Lambda^4 (1 - 2/r)}.
\eea
In the equatorial plane the line element is given by the relation
\be
  \d\tilde{l}^2 = \frac{\d r^2}{\left(1- 2/r\right)^2} +
    \frac{r^2}{(1+B^2r^2)^4 \left(1-2/r \right)}\,\d\phi^2.
\ee
This have to be identified with the line element (\ref{euclid}). We identify
the azimuthal coordinates again, and for the radial coordinates we find the
formulae
\be
  \rho = \frac{r}{(1+B^2r^2)^2 \sqrt{1-2/r}} \fcomma
\ee
and
\be
  \oder{\rho}{r} = \frac{1 - 3/r +B^2 r(5 - 3r)}{(1 + B^2r^2)^3
    \left(1 - 2/r\right)^{3/2}}.
\ee
We can see that also for the optical geometry of the Ernst spacetime, the
embedding diagrams shrink to zero radius asymptotically. Further, there is
$\d\rho/\d r = 0$ just at the loci of the circular photon orbits. Of course,
it is simply given by the fact that
\be
  \tilde{g}_{\phi\phi} = l_{\rm R}^2
\ee
at the equatorial plane. (Note that this is a general property of spherically
symmetric spacetimes \cite{SH}.) The embedding formula can be effectively
treated in the parametric form:
\be
  \left(\oder{z}{r}\right)^2 =
    \frac{\left(1-2/r\right)
    (1+B^2r^2)^6 -\left[1-3/r+B^2r(5-3r)\right]^2}%
    {\left(1-2/r\right)^3 (1+B^2r^2)^6}.
\ee
The embedding condition
\be                                                           \label{embedcond} 
  \left(1-\frac{2}{r}\right) (1+B^2r^2)^6 -
  \left[1-\frac{3}{r}+B^2r(5-3r)\right]^2 \geq 0
\ee
%is now very complicated, and can be treated by a numerical code only.
can be treated numerically only.
For $B=0$ it is reduced to the well-known Schwarzschild condition $r>9/4$.
%The results of the numerical code,
The numerically obtained results
giving the radii of the limit of embeddability $r_{\rm e}$, are illustrated in
Fig.\ \ref{f1}. The embedding diagrams are drawn in Fig.\ \ref{f4} for some
typical values of the parameter $B$, emphasizing qualitatively different
behaviour. For comparison with the Schwarzschild case, we give the functions
$z(\rho)$ in Fig.\ \ref{f6}.  Notice that the turning points of the embedding
diagrams correspond just to the circular photon orbits.

In Fig.\ \ref{f7} we give the radii $r_{\rm i}$, $r_{\rm o}$, $r_{\rm max}$,
and $r_{\rm e}$ in terms of the Euclidean radial coordinate $\rho$. We can see
that the outer turning point of the embedding diagrams, corresponding to
$r_{\rm o}$, is located behind the limit of embeddability, corresponding to
$r_{\rm e}$, if $B < B_{\rm o} \sim 0.055 < B_{\rm c}$.

\section{Concluding remarks}  \label{concl}

We have demonstrated through the analysis of the structure of the photon
capture cones and the embedding diagrams of both the ordinary and optical
geometry that the Ernst spacetimes have some highly different properties in
comparison with Schwarzschild spacetime.
The global properties of the Ernst spacetimes are imprinted in the character
of the photon capture cones of static observers at each radius $r>2$. Namely,
the asymptotic structure of the Ernst spacetimes reflects itself in the fact
that at any radius above the event horizon only the purely radial outward
directed photons can escape to infinity. This fact, together with the
existence of captured photons at $r_{\rm i} < r < r_{\rm o}$, makes the
structure of the photon capture cones in the Ernst spacetimes to be completely
different from the limiting case of the Schwarzschild spacetime at any radius
above the event horizon.

% BEGIN NEW SUBVERSION 1
%On the other hand, since the Ernst spacetimes reduce to the Schwarzschild
%spacetime for $B\rightarrow 0$, one can expect intuitively that for weak
%magnetic fields ($B\ll 1$) there could be some regions in the vicinity of the
%black-hole horizon, where gravitational (and inertial) phenomena determining,
%e.g., structure of an accretion disk, have similar character as in the
%Schwarzschild spacetimes. The phenomena are reflected by the embedding
%diagrams (see \cite{MTW}). Thus, we can conclude from the character of the
%embedding diagrams of both the ordinary and optical geometry, namely from the
%manner in which the Ernst embeddings peel off the Schwarzschild embedding (see
%Figs \ref{f5} and \ref{f6}), that such `regions of similarity' can be
%considered only for the Ernst spacetimes with $B\ll 1$, and at regions
%$r<1/B$. Of course, the concept of `similarity' should be put into a precise
%statement. We plan for a future work to do it by considering tidal effects in
%the Ernst spacetimes.
% END NEW SUBVERSION 1

% BEGIN NEW SUBVERSION 2
Moreover, the character of the embedding diagrams of both the ordinary and
optical geometry (Figs \ref{f5} and \ref{f6}) reflecting locally the
gravitational and inertial phenomena (see, e.g., \cite{MTW}), indicates that
these phenomena have to be qualitatively different from the Schwarzschild ones
at all $r>2$ for $B\sim 1$, and at $r>1/B$ for $B\ll 1$.
% END NEW SUBVERSION 2
% END NEW VERSION

\ack

The present work was supported by the GA\v{C}R grant No. 202/96/0206 and by
the Abdus Salam International Centre for Theoretical Physics in Trieste. Its
important part has been done during the visit of the authors at the ICTP\@.
The authors would like to thank Prof.\ M. A. Abramowicz, Prof.\ D. Sciama and
the Head of the High Energy Sector, Prof.\ S. Randjbar-Daemi for kind
hospitality at the ICTP.

\end{document}